\pdfoutput=1
\documentclass[12pt,a4paper]{article}

\usepackage{ifthen} 
\newboolean{pdflatex}
\setboolean{pdflatex}{true} 

\newboolean{articletitles}
\setboolean{articletitles}{true} 

\newboolean{uprightparticles}
\setboolean{uprightparticles}{false} 

\def\paperauthors{LHCb collaboration} 

\def\paperasciititle{First observation of the Lb->LcKKpi decay}

\def\papertitle{Observation of the \LbLckkpi decay} 

\def\paperkeywords{{High Energy Physics}, {LHCb}} 
\def\papercopyright{CERN on behalf of the LHCb collaboration}

\def\papercopyright{\the\year\ CERN for the benefit of the LHCb collaboration} 
\def\paperlicence{CC BY 4.0 licence}
\def\paperlicenceurl{https://creativecommons.org/licenses/by/4.0/}

\usepackage{longtable} 
\usepackage{subcaption} 

\usepackage[top=1in, bottom=1.25in, left=1in, right=1in]{geometry}

\usepackage{microtype}
\usepackage{lineno}  
\usepackage{xspace} 
\usepackage{caption} 

\usepackage{graphicx}  
\usepackage{color}
\usepackage{colortbl}
\graphicspath{{./figs/}} 

\usepackage{amsmath} 
\usepackage{amssymb}
\usepackage{amsfonts}
\usepackage{upgreek} 

\newcommand*\patchAmsMathEnvironmentForLineno[1]{%
\expandafter\let\csname old#1\expandafter\endcsname\csname #1\endcsname
\expandafter\let\csname oldend#1\expandafter\endcsname\csname
end#1\endcsname
 \renewenvironment{#1}%
   {\linenomath\csname old#1\endcsname}%
   {\csname oldend#1\endcsname\endlinenomath}%
}
\newcommand*\patchBothAmsMathEnvironmentsForLineno[1]{%
  \patchAmsMathEnvironmentForLineno{#1}%
  \patchAmsMathEnvironmentForLineno{#1*}%
}
\AtBeginDocument{%
\patchBothAmsMathEnvironmentsForLineno{equation}%
\patchBothAmsMathEnvironmentsForLineno{align}%
\patchBothAmsMathEnvironmentsForLineno{flalign}%
\patchBothAmsMathEnvironmentsForLineno{alignat}%
\patchBothAmsMathEnvironmentsForLineno{gather}%
\patchBothAmsMathEnvironmentsForLineno{multline}%
\patchBothAmsMathEnvironmentsForLineno{eqnarray}%
}

\usepackage{hyperxmp}

\usepackage[pdftex,
            pdfauthor={\paperauthors},
            pdftitle={\paperasciititle},
            pdfkeywords={\paperkeywords},
            pdfcopyright={Copyright (C) \papercopyright},
            pdflicenseurl={\paperlicenceurl}]{hyperref}

\usepackage[colorinlistoftodos,textsize=scriptsize]{todonotes}

\usepackage[bottom,flushmargin,hang,multiple]{footmisc}

\usepackage[all]{hypcap} 

\usepackage{mciteplus}
\usepackage{cite} 

\usepackage{xspace} 
\usepackage{upgreek}

\def\lhcb   {\mbox{LHCb}\xspace}

\def\rich   {RICH\xspace}

\def\hcal   {HCAL\xspace}
\def\MagUp {\mbox{\em Mag\kern -0.05em Up}\xspace}

\def\lone   {L0\xspace}

\ifthenelse{\boolean{uprightparticles}}%
{

 \def\Ppi         {\ensuremath{\uppi}\xspace}

 \def\Pchi        {\ensuremath{\upchi}\xspace}
 \def\Ppsi        {\ensuremath{\uppsi}\xspace}

 \def\PDelta      {\ensuremath{\Delta}\xspace}
 \def\PXi         {\ensuremath{\Xi}\xspace}
 \def\PLambda     {\ensuremath{\Lambda}\xspace}
 \def\PSigma      {\ensuremath{\Sigma}\xspace}
 \def\POmega      {\ensuremath{\Omega}\xspace}
 \def\PUpsilon    {\ensuremath{\Upsilon}\xspace}

 \def\PB      {\ensuremath{\mathrm{B}}\xspace}
 
 \def\PD      {\ensuremath{\mathrm{D}}\xspace}

 \def\PJ      {\ensuremath{\mathrm{J}}\xspace}
 \def\PK      {\ensuremath{\mathrm{K}}\xspace}

 \def\Pb      {\ensuremath{\mathrm{b}}\xspace}
 \def\Pc      {\ensuremath{\mathrm{c}}\xspace}

 \def\Pi      {\ensuremath{\mathrm{i}}\xspace}

 \def\Pp      {\ensuremath{\mathrm{p}}\xspace}

 \def\Ps      {\ensuremath{\mathrm{s}}\xspace}

 \def\thebaroffset{0.0em}
 }
 {

 \def\Ppi         {\ensuremath{\pi}\xspace}

 \def\Pchi        {\ensuremath{\chi}\xspace}
 \def\Ppsi        {\ensuremath{\psi}\xspace}
 
 \mathchardef\PDelta="7101
 \mathchardef\PXi="7104
 \mathchardef\PLambda="7103
 \mathchardef\PSigma="7106
 \mathchardef\POmega="710A
 \mathchardef\PUpsilon="7107

 \def\PB      {\ensuremath{B}\xspace}
 
 \def\PD      {\ensuremath{D}\xspace}

 \def\PJ      {\ensuremath{J}\xspace}
 \def\PK      {\ensuremath{K}\xspace}

 \def\Pb      {\ensuremath{b}\xspace}
 \def\Pc      {\ensuremath{c}\xspace}

 \def\Pi      {\ensuremath{i}\xspace}

 \def\Pp      {\ensuremath{p}\xspace}

 \def\Ps      {\ensuremath{s}\xspace}

 \def\thebaroffset{0.18em}
}
\newcommand{\offsetoverline}[2][\thebaroffset]{\kern #1\overline{\kern -#1 #2}}%

\makeatletter
\ifcase \@ptsize \relax
  \newcommand{\miniscule}{\@setfontsize\miniscule{4}{5}}
\or
  \newcommand{\miniscule}{\@setfontsize\miniscule{5}{6}}
\or
  \newcommand{\miniscule}{\@setfontsize\miniscule{5}{6}}
\fi
\makeatother

\DeclareRobustCommand{\optbar}[1]{\shortstack{{\miniscule (\rule[.5ex]{1.25em}{.18mm})}
  \\ [-.7ex] $#1$}}

\def\squark    {{\ensuremath{\Ps}}\xspace}

\def\cquark    {{\ensuremath{\Pc}}\xspace}

\def\bquark    {{\ensuremath{\Pb}}\xspace}

\def\pion   {{\ensuremath{\Ppi}}\xspace}
\def\piz    {{\ensuremath{\pion^0}}\xspace}
\def\pip    {{\ensuremath{\pion^+}}\xspace}
\def\pim    {{\ensuremath{\pion^-}}\xspace}

\def\kaon    {{\ensuremath{\PK}}\xspace}

\def\KorKbar {\kern \thebaroffset\optbar{\kern -\thebaroffset \PK}{}\xspace}

\def\Kp      {{\ensuremath{\kaon^+}}\xspace}
\def\Km      {{\ensuremath{\kaon^-}}\xspace}

\def\Dbar    {{\ensuremath{\offsetoverline{\PD}}}\xspace}
\def\D       {{\ensuremath{\PD}}\xspace}

\def\DorDbar {\kern \thebaroffset\optbar{\kern -\thebaroffset \PD}\xspace}
\def\Dz      {{\ensuremath{\D^0}}\xspace}
\def\Dzb     {{\ensuremath{\Dbar{}^0}}\xspace}
\def\Dp      {{\ensuremath{\D^+}}\xspace}
\def\Dm      {{\ensuremath{\D^-}}\xspace}

\def\DpDm    {\ensuremath{\Dp {\kern -0.16em \Dm}}\xspace}

\def\Dstarp  {{\ensuremath{\D^{*+}}}\xspace}

\def\Ds      {{\ensuremath{\D^+_\squark}}\xspace}

\def\Dsm     {{\ensuremath{\D^-_\squark}}\xspace}

\def\Dssm    {{\ensuremath{\D^{*-}_\squark}}\xspace}

\def\B       {{\ensuremath{\PB}}\xspace}
\def\Bbar    {{\ensuremath{\offsetoverline{\PB}}}\xspace}
\def\Bb      {{\ensuremath{\Bbar}}\xspace}
\def\BorBbar {\kern \thebaroffset\optbar{\kern -\thebaroffset \PB}\xspace}

\def\Bzb     {{\ensuremath{\Bbar{}^0}}\xspace}
\def\Bd      {{\ensuremath{\B^0}}\xspace}

\def\BdorBdbar {\kern \thebaroffset\optbar{\kern -\thebaroffset \Bd}\xspace}

\def\Bub     {{\ensuremath{\B^-}}\xspace}

\def\Bm      {{\ensuremath{\Bub}}\xspace}

\def\Bs      {{\ensuremath{\B^0_\squark}}\xspace}
\def\Bsb     {{\ensuremath{\Bbar{}^0_\squark}}\xspace}
\def\BsorBsbar {\kern \thebaroffset\optbar{\kern -\thebaroffset \Bs}\xspace}

\def\jpsi     {{\ensuremath{{\PJ\mskip -3mu/\mskip -2mu\Ppsi\mskip 2mu}}}\xspace}
\def\psitwos  {{\ensuremath{\Ppsi{(2S)}}}\xspace}

\def\chicone  {{\ensuremath{\Pchi_{\cquark 1}}}\xspace}
\def\chictwo  {{\ensuremath{\Pchi_{\cquark 2}}}\xspace}

\def\Y#1S{\ensuremath{\PUpsilon{(#1S)}}\xspace}

\def\proton      {{\ensuremath{\Pp}}\xspace}
\def\antiproton  {{\ensuremath{\overline \proton}}\xspace}

\def\Lz          {{\ensuremath{\PLambda}}\xspace}

\def\LorLbar     {\kern \thebaroffset\optbar{\kern -\thebaroffset \PLambda}\xspace}

\def\Sigmares    {{\ensuremath{\PSigma}}\xspace}

\def\Xires       {{\ensuremath{\PXi}}\xspace}

\def\Omegares    {{\ensuremath{\POmega}}\xspace}

\def\Lc          {{\ensuremath{\Lz^+_\cquark}}\xspace}

\def\Lb           {{\ensuremath{\Lz^0_\bquark}}\xspace}

\def\Xib          {{\ensuremath{\Xires_\bquark}}\xspace}

\def\Omegab       {{\ensuremath{\Omegares^-_\bquark}}\xspace}

\def\BF         {{\ensuremath{\mathcal{B}}}\xspace}
\def\BR         {\BF}

\newcommand{\decay}[2]{\ensuremath{#1\!\to #2}\xspace}

\def\to                 {\ensuremath{\rightarrow}\xspace}

\def\AT#1     {\ensuremath{A_{\mathrm{T}}^{#1}}\xspace}           

\def\C#1      {\ensuremath{\mathcal{C}_{#1}}\xspace}                       
\def\Cp#1     {\ensuremath{\mathcal{C}_{#1}^{'}}\xspace}                    
\def\Ceff#1   {\ensuremath{\mathcal{C}_{#1}^{\mathrm{(eff)}}}\xspace}        
\def\Cpeff#1  {\ensuremath{\mathcal{C}_{#1}^{'\mathrm{(eff)}}}\xspace}       
\def\Ope#1    {\ensuremath{\mathcal{O}_{#1}}\xspace}                       
\def\Opep#1   {\ensuremath{\mathcal{O}_{#1}^{'}}\xspace}                    

\newcommand{\nospaceunit}[1]{\ensuremath{\text{#1}}}
\newcommand{\aunit}[1]{\ensuremath{\text{\,#1}}}

\newcommand{\tev}{\aunit{Te\kern -0.1em V}\xspace}
\newcommand{\gev}{\aunit{Ge\kern -0.1em V}\xspace}
\newcommand{\mev}{\aunit{Me\kern -0.1em V}\xspace}
\newcommand{\kev}{\aunit{ke\kern -0.1em V}\xspace}
\newcommand{\ev}{\aunit{e\kern -0.1em V}\xspace}
\newcommand{\mevc}{\ensuremath{\aunit{Me\kern -0.1em V\!/}c}\xspace}
\newcommand{\gevc}{\ensuremath{\aunit{Ge\kern -0.1em V\!/}c}\xspace}
\newcommand{\mevcc}{\ensuremath{\aunit{Me\kern -0.1em V\!/}c^2}\xspace}
\newcommand{\gevcc}{\ensuremath{\aunit{Ge\kern -0.1em V\!/}c^2}\xspace}

\def\mum  {\ensuremath{\,\upmu\nospaceunit{m}}\xspace}

\def\fb   {\ensuremath{\aunit{fb}}\xspace}
\def\invfb   {\ensuremath{\fb^{-1}}\xspace}

\def\ps   {\ensuremath{\aunit{ps}}\xspace}

\newcommand{\chisq}{\ensuremath{\chi^2}\xspace}

\newcommand{\chisqip}{\ensuremath{\chi^2_{\text{IP}}}\xspace}

\newcommand{\chisqvtx}{\ensuremath{\chi^2_{\text{vtx}}}\xspace}

\def\gsim{{~\raise.15em\hbox{$>$}\kern-.85em
          \lower.35em\hbox{$\sim$}~}\xspace}
\def\lsim{{~\raise.15em\hbox{$<$}\kern-.85em
          \lower.35em\hbox{$\sim$}~}\xspace}

\def\sqs   {\ensuremath{\protect\sqrt{s}}\xspace}

\def\pt         {\ensuremath{p_{\mathrm{T}}}\xspace}

\def\ptot       {\ensuremath{p}\xspace}

\def\evtgen     {\mbox{\textsc{EvtGen}}\xspace}

\def\geant      {\mbox{\textsc{Geant4}}\xspace}

\def\photos     {\mbox{\textsc{Photos}}\xspace}

\def\pythia     {\mbox{\textsc{Pythia}}\xspace}

\def\tell1  {TELL1\xspace}
\def\ukl1   {UKL1\xspace}

\newcommand{\eg}{\mbox{\itshape e.g.}\xspace}
\newcommand{\ie}{\mbox{\itshape i.e.}\xspace}

\newcommand{\LbLckkpi}{\ensuremath{\Lb\to\Lc\Kp\Km\pim}\xspace}
\newcommand{\LbLcDs}{\ensuremath{\Lb\to\Lc\Dsm}\xspace}

\newcommand{\Dskkpi}{\ensuremath{\Dsm\to\Kp\Km\pim}\xspace}

\def\Sigmacp{{\ensuremath{\Sigmares_\cquark^+}}\xspace}

\begin{document}

\renewcommand{\thefootnote}{\fnsymbol{footnote}}
\setcounter{footnote}{1}

\begin{titlepage}
\pagenumbering{roman}

\vspace*{-1.5cm}
\centerline{\large EUROPEAN ORGANIZATION FOR NUCLEAR RESEARCH (CERN)}
\vspace*{1.5cm}
\noindent
\begin{tabular*}{\linewidth}{lc@{\extracolsep{\fill}}r@{\extracolsep{0pt}}}
\ifthenelse{\boolean{pdflatex}}
{\vspace*{-1.5cm}\mbox{\!\!\!\includegraphics[width=.14\textwidth]{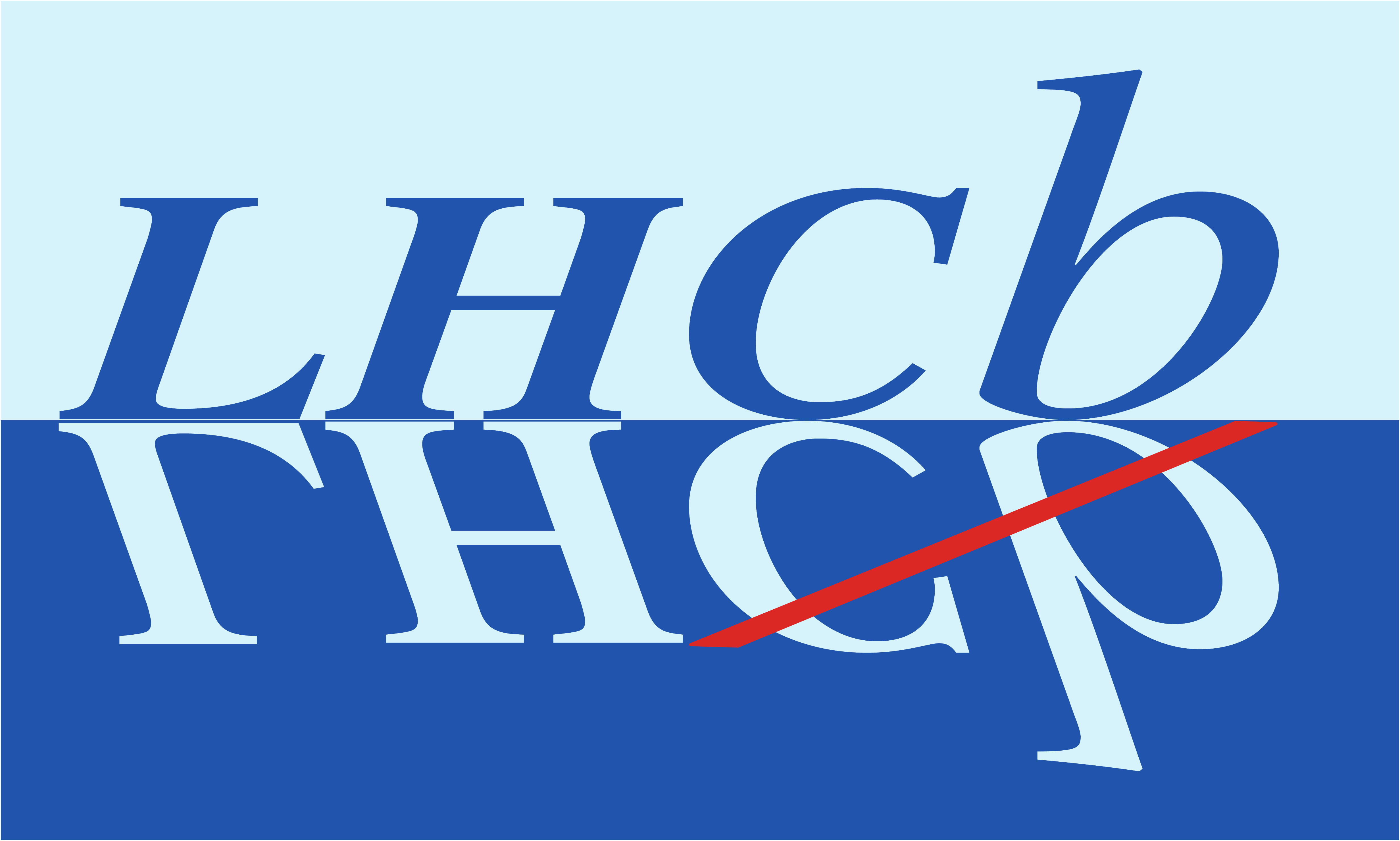}} & &}%
{\vspace*{-1.2cm}\mbox{\!\!\!\includegraphics[width=.12\textwidth]{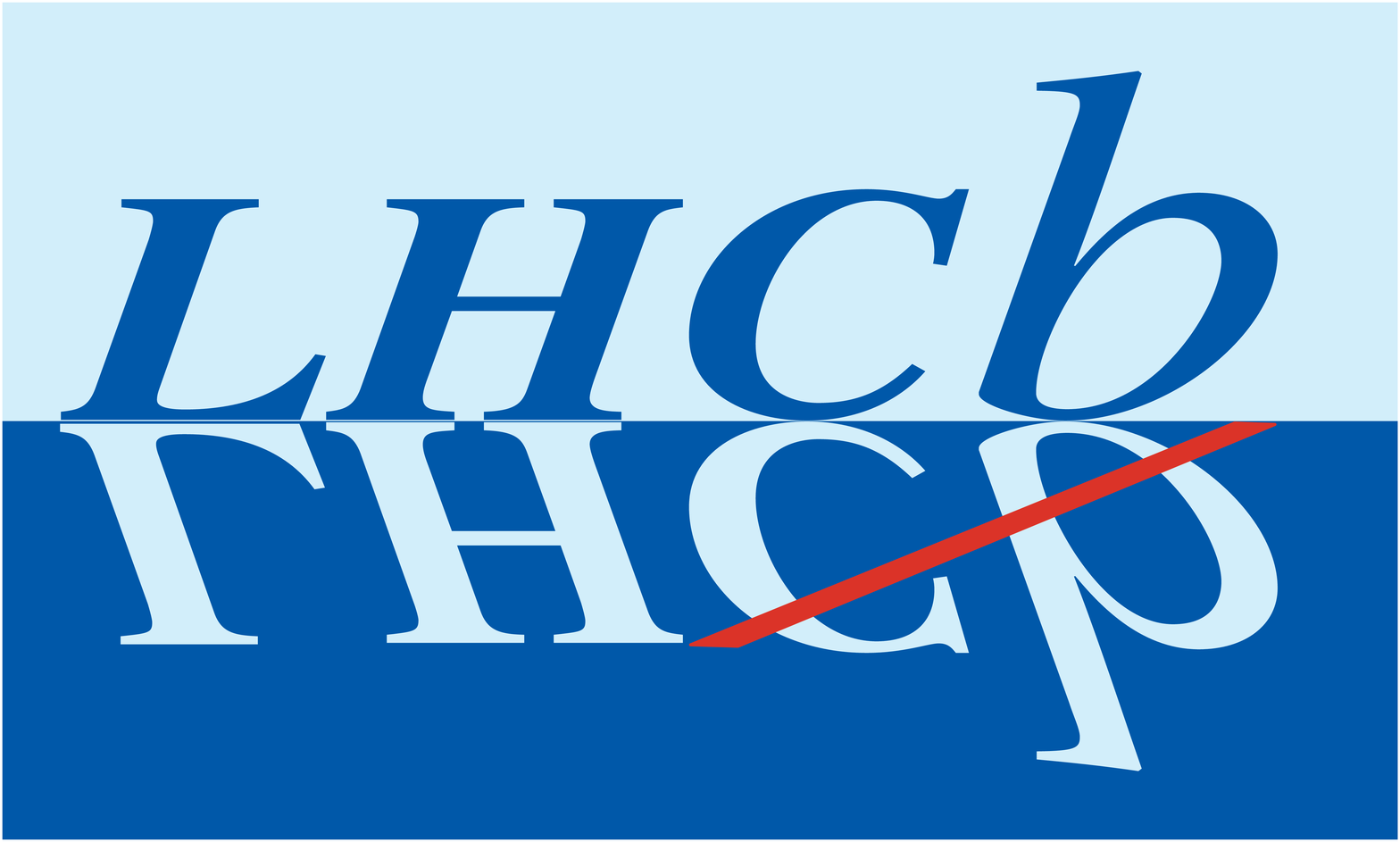}} & &}%
\\
 & & CERN-EP-2020-209 \\  
 & & LHCb-PAPER-2020-028 \\  
 & & November 28, 2020   \\               
 & & \\
\end{tabular*}

\vspace*{4.0cm}

{\normalfont\bfseries\boldmath\huge
\begin{center}
  \papertitle
\end{center}
}

\vspace*{2.0cm}

\begin{center}
\paperauthors\footnote{Authors are listed at the end of this paper.}
\end{center}

\vspace{\fill}

\begin{abstract}
  \noindent
   The $\varLambda^{0}_{b}\to{\varLambda^{+}_{c}\xspace}K^{+}K^{-}\pi^{-}$ decay is observed for the first time using a data sample of proton-proton collisions at centre-of-mass energies of $\ensuremath{\protect\sqrt{s}}\xspace=7$ and 8 $\rm{{Te\kern -0.1em V}}\xspace$  collected by the {\mbox{LHCb}\xspace} detector,
   corresponding to an integrated luminosity of $3\xspace\ensuremath{\rm{fb}\xspace^{-1}}\xspace$.
   The ratio of branching fractions between the ${{\ensuremath{\varLambda^{0}_{b}}}\xspace}\to{{\ensuremath{\varLambda^{+}_{c}}}\xspace}K^{+}K^{-}\pi^{-}$
   and the $\varLambda^{0}_{b}\to\varLambda^{+}_{c} D^{-}_{s}$ decays is measured to be
\begin{equation*}
        \frac{{\mathcal{B}}(\varLambda^{0}_{b}\to{\varLambda^{+}_{c}\xspace}K^{+}K^{-}\pi^{-}) }
        {{\mathcal{B}}( \varLambda^{0}_{b}\to\varLambda^{+}_{c} D^{-}_{s})}
        =  (9.26 \pm 0.29 \pm 0.46 \pm 0.26)\times10^{-2},
\end{equation*}
   where the first uncertainty is statistical, the second systematic and the third is due to the knowledge of the $D^{-}_{s}\to K^{+}K^{-}\pi^{-}$ branching fraction.
   No structure on the invariant mass distribution of the $\varLambda^{+}_{c}K^{+}$ system is found, consistent with no open-charm pentaquark signature.
\end{abstract}


\vspace*{2.0cm}

\begin{center}
  Published in 
   Phys.~Lett.~B 815 (2021) 136172
\end{center}

\vspace{\fill}

{\footnotesize
\centerline{\copyright~\papercopyright. \href{\paperlicenceurl}{\paperlicence}.}}
\vspace*{2mm}

\end{titlepage}


\newpage
\setcounter{page}{2}
\mbox{~}
%

\renewcommand{\thefootnote}{\arabic{footnote}}
\setcounter{footnote}{0}

\cleardoublepage

\pagestyle{plain} 
\setcounter{page}{1}
\pagenumbering{arabic}


\section{Introduction}
\label{sec:Introduction}

Over the last two decades, a wealth of information has been accumulated on the decays of hadrons containing \bquark quarks~\cite{PDG2020}.
Measurements of their decay rates and properties have been used to test the Cabibbo-Kobayashi-Maskawa  mechanism~\cite{PhysRevLett.10.531,Kobayashi:1973fv} describing weak interactions within the Standard Model, and to examine various theoretical approaches, 
such as the heavy quark effective theory~\cite{EICHTEN1990511} and the factorization hypothesis~\cite{Fakirov:1977ta,Cabibbo:1977zv,MANNEL1991359,BENEKE2000313}.
Although many \bquark-hadron decays have been observed with their branching fractions measured, a large number of them remains either unobserved or poorly measured, most notably decays of \Lb, \Xib and \Omegab baryons. In the last years, the LHCb experiment has observed many new $\Lb$ decays to final states  
such as $\Lc\pim\pip\pim$~\cite{LHCb-PAPER-2011-016}, $\Lc \pim \proton \antiproton$~\cite{LHCb-PAPER-2018-005}, $\Lc \Dsm$~\cite{LHCb-PAPER-2014-002}, 
$\chicone\proton\Km$, $\chictwo\proton\Km$~\cite{LHCb-PAPER-2017-011},
$\psitwos\proton\Km$ and $\jpsi\pip\pim\proton\Km$~\cite{LHCb-PAPER-2015-060}.\footnote{The charge-conjugate process is implied throughout this Letter.} 

In this Letter, the first observation of the \LbLckkpi decay (referred to hereafter as signal channel) is reported, along with a measurement of its branching fraction relative to that of the \LbLcDs decay (normalisation channel). The analysis
uses a data sample of proton-proton ($pp$) collisions at centre-of-mass energies of $\sqs=7$ and 8\tev collected by the \lhcb experiment, corresponding to an integrated luminosity of $3\invfb$. The observation of the $\LbLckkpi$ decay provides a laboratory to search for open-charm pentaquarks with valence quark content $c\bar{s}uud$ that could decay strongly to the $\Lc\Kp$ final state.  These states are a natural extension of the three narrow pentaquark candidates with quark content $c\bar{c}uud$ observed in $\Lb\to\jpsi p \Km$ decays~\cite{LHCb-PAPER-2019-014}, with the $\bar{c}$ quark replaced by an $\bar{s}$ quark. 
The recent discovery of a $\Dp\Km$ structure in $\Bm\to \Dm \Dp\Km$ decays~\cite{LHCb-PAPER-2020-024,LHCb-PAPER-2020-025}, consistent with open-charm tetraquarks,  also motivates the search for open-charm pentaquarks.  

\begin{figure}[h ]
\begin{center}
\includegraphics[width=0.6\linewidth]{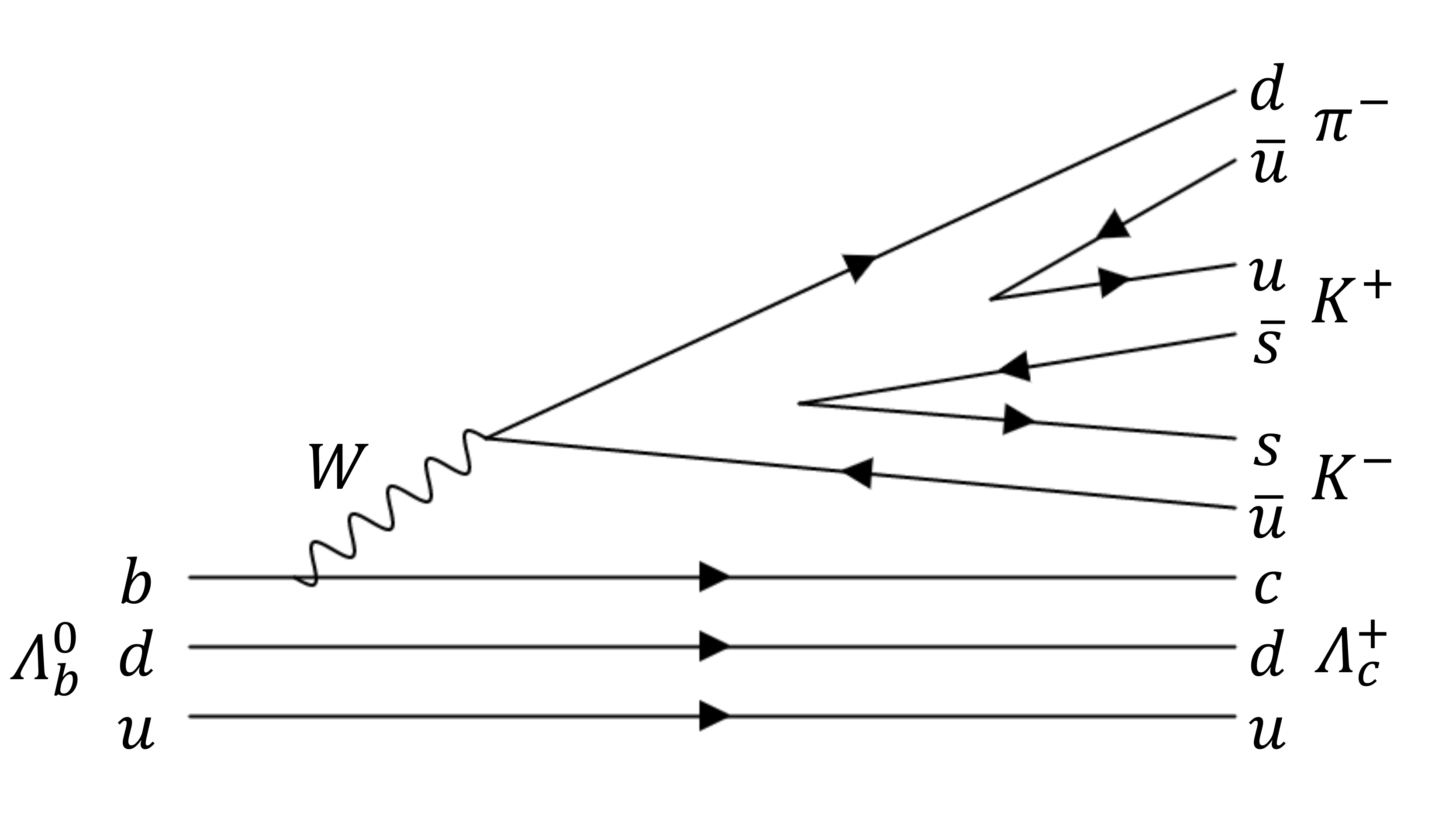}
\end{center}
  \caption{ \small{Feynman diagram of the leading contribution to the \LbLckkpi signal decay.}  }
  \label{fig:Feynman}
\end{figure}

Figure~\ref{fig:Feynman} shows the leading diagram contributing to the signal decay. Contributions to the companion $K^+K^-\pi^-$ system could be through intermediate  $a_1^-$ mesons, such as the $a_1(1260)^-$ state, which is found to dominate in $\Bb\to D^{(*)} K^{*0}K^-$ decays~\cite{Drutskoy:2002ib}. Decays of $\PSigma_c^0\to \Lc \pi^-$ or even $\PXi_c^0\to \Lc K^-$ could also contribute to the signal.

\section{Detector and simulation}
\label{sec:Detector}

The \lhcb detector~\cite{LHCb-DP-2008-001,LHCb-DP-2014-002} is a single-arm forward
spectrometer covering the \mbox{pseudorapidity} range $2<\eta <5$,
designed for the study of particles containing \bquark or \cquark
quarks. The detector includes a high-precision tracking system
consisting of a silicon-strip vertex (VELO) detector surrounding the $pp$
interaction region, a large-area silicon-strip detector located
upstream of a dipole magnet with a bending power of about
$4{\mathrm{\,Tm}}$, and three stations of silicon-strip detectors and straw
drift tubes placed downstream of the magnet.
The tracking system provides a measurement of the momentum, \ptot, of charged particles with
a relative uncertainty that varies from 0.5\% at low momentum to 1.0\% at 200\gevc.
The minimum distance of a track to a primary $pp$ collision vertex (PV), the impact parameter (IP), 
is measured with a resolution of $(15+29/\pt)\mum$,
where \pt is the component of the momentum transverse to the beam, in\,\gevc.
Different types of charged hadrons are distinguished using information
from two ring-imaging Cherenkov detectors. 
Photons, electrons and hadrons are identified by a calorimeter system consisting of
scintillating-pad and preshower detectors, an electromagnetic
and a hadronic calorimeter (HCAL). Muons are identified by a
system composed of alternating layers of iron and multiwire proportional chambers.

The online event selection is performed by a trigger based on signal information only.
The trigger consists of a hardware stage, based on information from the calorimeter system, followed by a software stage, which applies a full event
reconstruction~\cite{LHCb-DP-2012-004}.
At the hardware trigger stage, 
referred to as L0 trigger in the following,
the \decay{\Lb}{\Lc\Kp\Km\pim} and \decay{\Lb}{\Lc\Dsm} candidates are required to include a hadron having high transverse energy deposited in the calorimeters. The transverse energy threshold is 3.5\gev.
The software trigger, also named high-level trigger (HLT), requires a two-, three- or four-track vertex with a significant displacement from any PV. At least one charged particle
must have a large transverse momentum and be
inconsistent with originating from any PV. 
A multivariate algorithm~\cite{BBDT} is used for 
the identification of displaced vertices consistent with the decay
of a \bquark-hadron.

Simulation is used to model the effects of the detector acceptance and the 
selection requirements, to validate the fit models and to evaluate 
efficiencies. 
In the simulation, $pp$ collisions are generated using
\pythia~8~\cite{Sjostrand:2007gs} with a specific \lhcb
configuration~\cite{LHCb-PROC-2010-056}.  
Decays of unstable particles
are described by \evtgen~\cite{Lange:2001uf}, in which final-state
radiation is generated using \photos~\cite{Golonka:2005pn}. The 
interaction of the generated particles with the detector, and its response,
are implemented using the \geant
toolkit~\cite{Allison:2006ve, *Agostinelli:2002hh} as described in
Ref.~\cite{LHCb-PROC-2011-006}.

\section{Event selection}

Candidate \Lc and \Dsm hadrons are reconstructed through their decays to the $\proton\Km\pip$ and $\Kp\Km\pim$ final states, respectively.
The offline candidate selection is performed by applying a loose preselection, followed by a multivariate analysis (MVA) to further suppress  combinatorial background originating from  random combinations.
To reduce systematic uncertainties on the ratio of efficiencies between the signal and the normalisation channels,
the selection criteria of \Lc candidates are identical between the two channels.

A good-quality track with $\pt>100\mevc$ and $p>1\gevc$
is required for each final-state particle.
Protons and antiprotons are required to have a momentum greater than $10\gevc$
to improve their identification.
All final-state particles are also required to be inconsistent with originating from any PV by requiring a large \chisqip,
where \chisqip is defined as the difference in the \chisq of a given PV fit with and without the track under consideration.
Each \Lc baryon candidate is required to have at least one decay product with $\pt>500\mevc$ and $p>5\gevc$,
a good-quality vertex (\ie small \chisqvtx), and invariant mass within $\pm15\mevcc$
of the known \Lc mass\cite{PDG2020}.
For the \Lc candidates, the sum of transverse momenta of their decay products must exceed 1.8\gevc.
The selection criteria for \Dsm candidates are similar to
those of \Lc candidates.
The $\Kp\Km\pim$ invariant mass is required to be within $\pm35\mevcc$ from the known \Dsm meson mass.

The signal channel is reconstructed by combining
\Lc, \Kp, \Km and \pim candidates,
while the normalisation channel is reconstructed by combining a \Lc with a \Dsm candidate.
The combinations above form \Lb candidates, which are required to have
a small \chisqvtx and \chisqip,
and a decay time with respect to its associated PV greater than 0.2\ps.
The associated PV is the one that gives the smallest \chisqip, 
where the \chisqip denotes the IP significance of candidate’s trajectory returned by the kinematical fit. 
The angle between the $\Lb$ momentum and the vector pointing from the associated PV to the $\Lb$ decay vertex, $\theta_p$,  is required to be smaller than 11 mrad.
The \Lb candidate is also required to have at least one final-state particle with $\pt>1.7\gevc$, 
and its decay vertex significantly displaced from any PV.
The latter is achieved by requiring the significance of the flight distance between the \Lb decay vertex and any PV to be larger than 4.
Final-state tracks of signal and normalisation candidates
must pass stringent particle-identification requirements based on the information from \rich detectors, calorimeter system and muon stations.
To reject tracks that share the same segment in the VELO detector, 
any two tracks with the same charge used to form the $\Lb$ candidate are required to have an opening angle larger than 0.5 mrad.
A kinematic fit~\cite{Hulsbergen:2005pu} of the decay chain constrains the \Lb candidate
to originate from the associated PV and the \Lc candidate invariant mass to its known value~\cite{PDG2020}.

The $\Lb$ candidate could originate from $\Bzb\to \Dp K^+K^-\pi^-$ or $\Bsb\to\Ds K^+K^-\pi^-$ decays, where a pion or kaon in $\Dp\to\Kp\pim\pip$ or $\Ds(\Dp)\to \Kp\Km\pip$ decays is misidentified as a proton.
These background contributions are vetoed if the invariant masses of the \Lc and \Lb candidates, evaluated by replacing the proton by either the pion or kaon mass hypothesis, are within $\pm15$\mevcc
of the known $\Dp(\Ds)$ mass and $\pm25$\mevcc of the known $\Bzb(\Bsb)$ mass~\cite{PDG2020}. These vetoes are applied to both the signal and the normalisation channels. For the signal decay, additional vetoes are applied if the invariant mass of the $K^+\pi^-$ or $K^+K^-\pim$ companion tracks falls within $\pm30$\mevcc of the $\Dzb$ or $\Dsm$ known mass, respectively~\cite{PDG2020}.

Reconstructed candidates are further required to pass an MVA output threshold based upon a multilayer perceptron (MLP) filter\cite{TMVA4}, designed to reject the combinatorial background.
The MLP classifier is trained using a signal sample of simulated \LbLckkpi decays tuned on data to reproduce correctly the \Lb production kinematics based on the $\pt$ and $y$ distributions 
and a background sample taken from the upper sideband of the \Lb invariant mass spectrum in the range of \mbox{5.75 -- 7\gevcc}. 
A four-body phase-space simulation is used for the signal sample to keep the MLP efficiency as uniform as possible,
as including intermediate resonances in the simulation could potentially lead to small MLP efficiencies for less represented phase-space regions.
The lower sideband is not used to avoid potential background contributions from partially reconstructed decays.
The MLP input includes the following variables:
\pt sum of the \Lc decay products,
minimal \chisqip among the \Lc decay products,
minimal \pt and minimal \chisqip among the kaons originating directly from the \Lb decay,
\pt and \chisqip of the \pim from the $\Lb$ decay,
\pt sum of all $\pi$ and $K$ originating directly from the $\Lb$ decay,
\chisqvtx of the \Lc candidate,
$\chisq$ of the flight distance between the $\Lb$ decay vertex and the associated PV,
$\cos\theta_p$,
$\chisq$ probability of the \Lb candidate vertex fit,
and the difference of longitudinal position between the \Lc and the \Lb decay vertices.


The MLP response obtained from the training is also applied to the normalisation channel sample. 
The optimal thresholds on the MLP response are obtained for the signal and normalisation channels separately by maximising a figure-of-merit, 
defined as $S/\sqrt{S+B}$,
where $S$ and $B$ are the expected signal and background yields for \Lb candidates within a $\pm2.5\,\sigma$ mass window around the known \Lb mass~\cite{PDG2020}, 
where $\sigma$ is the mass resolution corresponding to about 12\mevcc.
Both $S$ and $B$ are determined by multiplying the initial yields of signal 
and background with the corresponding MLP selection efficiencies estimated from simulation and sideband data, respectively. 
The initial signal and background yields are obtained from a preliminary fit to the preselected data sample before the MLP requirement applied, 
where the signal \Lb peak is already seen in the $\Lc\Kp\Km\pim$ invariant mass distribution.
The optimal point corresponds to a signal efficiency of 90\% and a background rejection of 85\%.
About 0.6\% events in the signal channel contain multiple candidates, 
only one candidate is retained by a random selection.

\section{Signal yields and search for intermediate states}
\label{sec:fit}
\begin{figure}[tb]
  \begin{center}
    \includegraphics[width=0.49\linewidth]{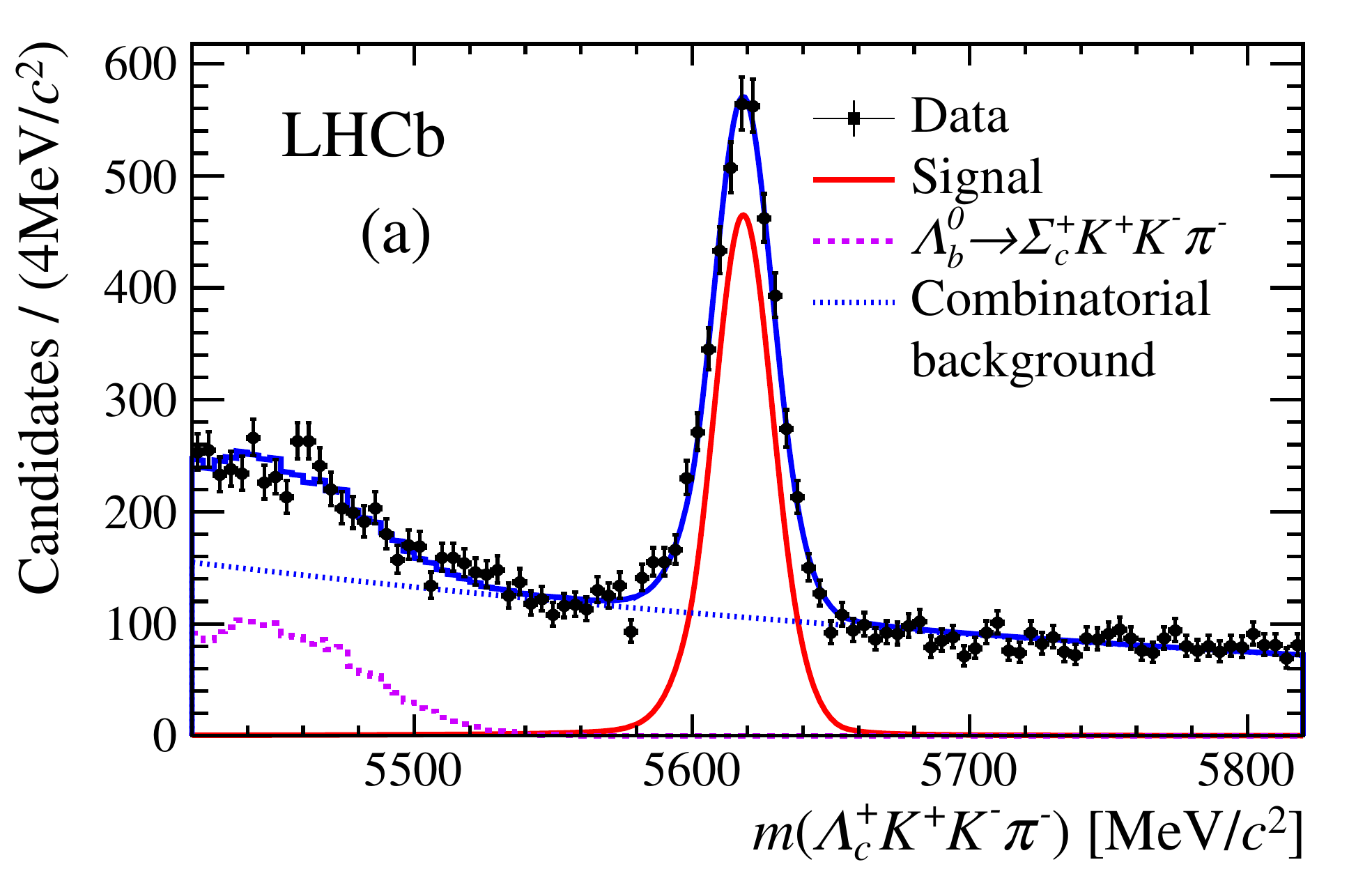}
    \includegraphics[width=0.49\linewidth]{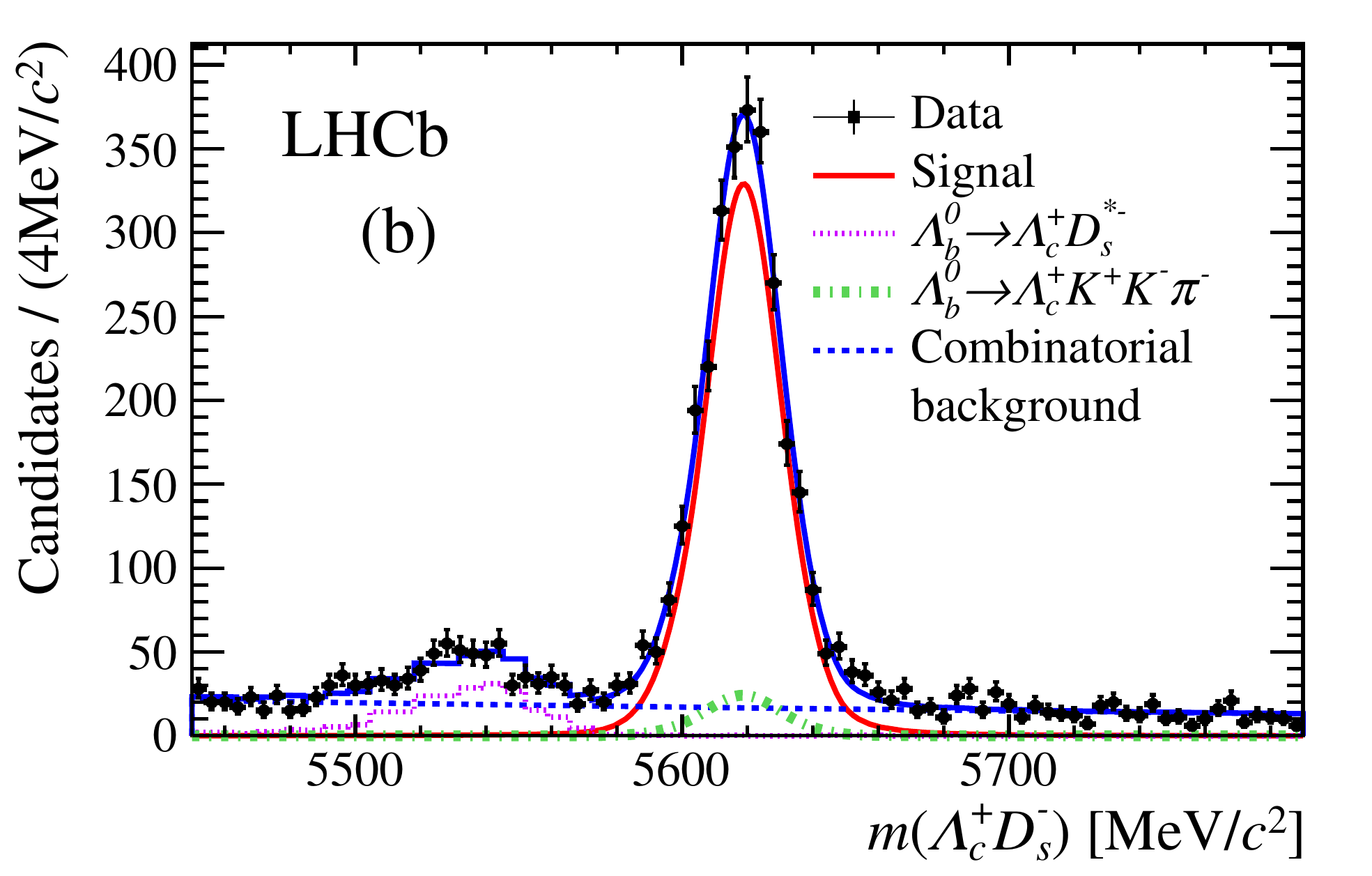}
    \vspace*{-0.5cm}
  \end{center}
  \caption{
    \small 
    Invariant mass distribution of (a) \LbLckkpi and (b) \LbLcDs candidates.
    Fit projections are overlaid as a blue solid line.
    For (a), 
    the red solid line represents the signal component, 
    the blue dashed line is the background due to random combinations,
    and the violet dotted line is the contribution from $\Lb\to\Sigmacp\Kp\Km\pim$ decays.
    For (b),
    the red solid line is the normalization channel component,
    the violet dotted line is the $\Lb\to\Lc\Dssm$ background,
    the green dashed-dotted line is the contribution from \LbLckkpi decays, 
    and the blue dashed line represents combinatorial background.
    }   
   \label{fig:fit}
\end{figure}

The yields in both the signal and  normalisation channels are determined 
from an unbinned extended maximum-likelihood fit to the corresponding invariant mass spectra of the \Lc\Kp\Km\pim system.
The signal component is modelled by a sum of two Crystal Ball functions \cite{Skwarnicki:1986xj} with a common mean of the Gaussian cores,
with tail parameters fixed to the values obtained from simulation.
For both the signal and normalisation channels, 
the combinatorial background is described by an exponential function,
whose parameters are varied freely and allowed to be different between the signal and normalisation channels.
For the signal channel, 
a significant contribution from $\Lb\to\Sigmacp[\to\Lc\piz]\Kp\Km\pim$ decays is present in the lower invariant mass region, 
which has the same final state as the $\piz$ is not reconstructed.
The shape of this background is obtained from a simulation of $\Lb\to\PSigma_c(2455)^+\Kp\Km\pim$ decays. 
For the normalisation channel, 
the $\Lb\to\Lc\Dssm$ decay may be reconstructed as $\Lb\to\Lc\Dsm$ due to photon emission in the \Dssm decay.
The shape of this background is obtained from simulated $\Lb\to\Lc\Dssm$  decays.
The signal decay can also contribute to the normalisation channel 
forming a background
under the $\Dsm$ mass peak. 
This background contribution is estimated from the \Dsm sidebands of the normalisation data sample, 
where the width of the sideband is chosen to be the same as that of the \Dsm mass window used in the normalisation channel selection.
The invariant mass distributions for the signal and normalisation channels 
are shown in Fig.~\ref{fig:fit} with the fit projects overlaid.
The signal yields are obtained to be $N(\LbLckkpi)=3\,400\pm80$ and $N(\Lb\to\Lc\Dsm[\Kp\Km\pim])=2\,550\pm60$, respectively, where the uncertainties are statistical only.

\begin{figure}[!tb]
  \begin{center}
    \includegraphics[width=0.49\linewidth]{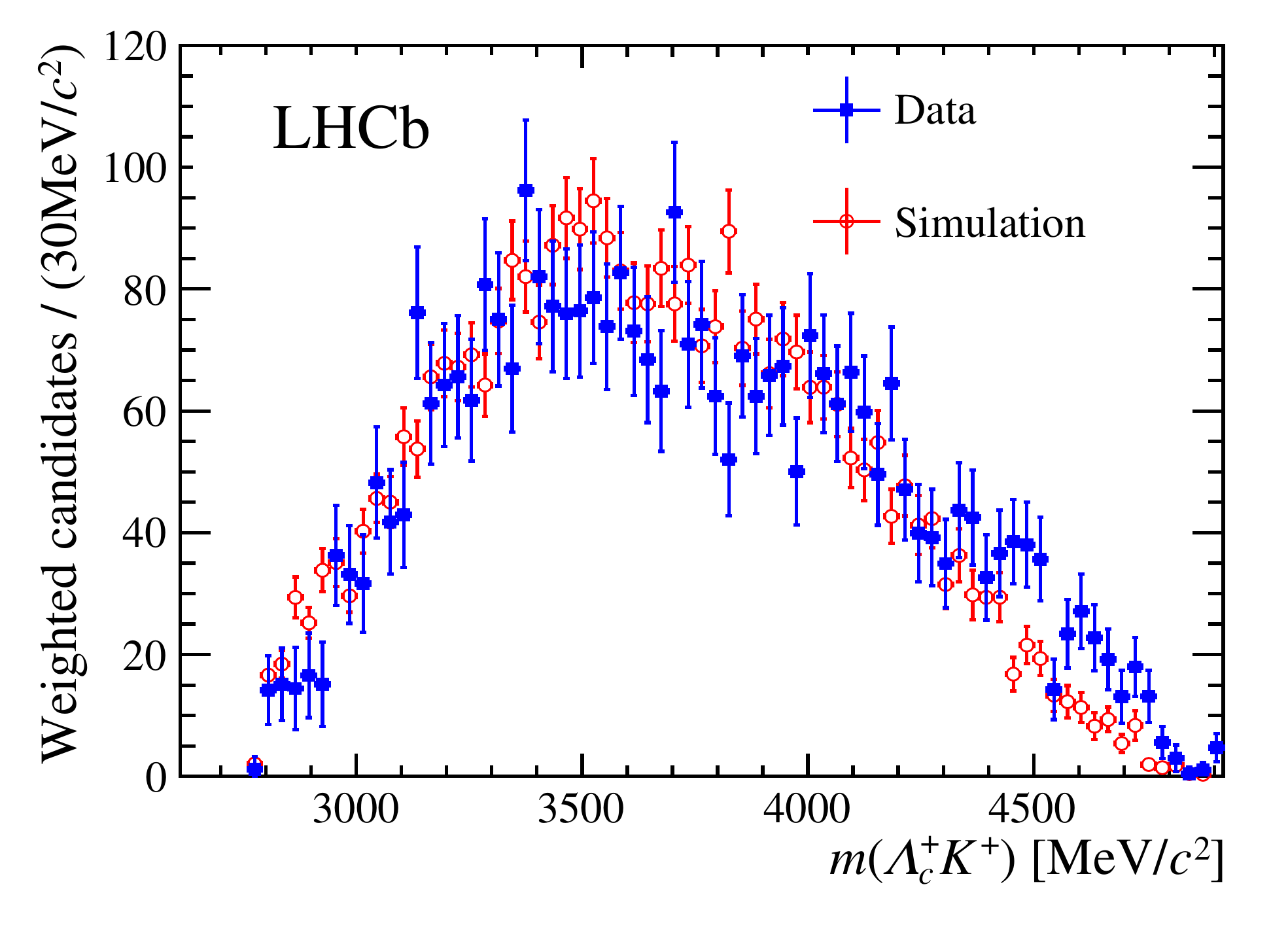}
    \vspace*{-0.5cm}
  \end{center}
  \caption{
    \small 
    Invariant mass distributions of \Lc\Kp candidates in the \LbLckkpi signal channel 
    for  the simulation (red line) and the background-subtracted data (blue points with error bars).
    }   
   \label{fig:dis}
\end{figure}

An open-charm pentaquark state could be revealed as a structure in the invariant mass distribution of the \Lc\Kp system, shown in Fig.~\ref{fig:dis} for data and simulation.
The data distribution is background subtracted through the {\it sPlot} weighting technique\cite{Pivk:2004ty}, using the \Lc\Kp\Km\pim invariant mass as discriminating variable. No structure is observed. A full amplitude analysis is needed to estimate the limit of the pentaquark contribution, which is beyond the scope of this Letter. 

\begin{figure}[tb]
  \begin{center}
    \includegraphics[width=0.49\textwidth]{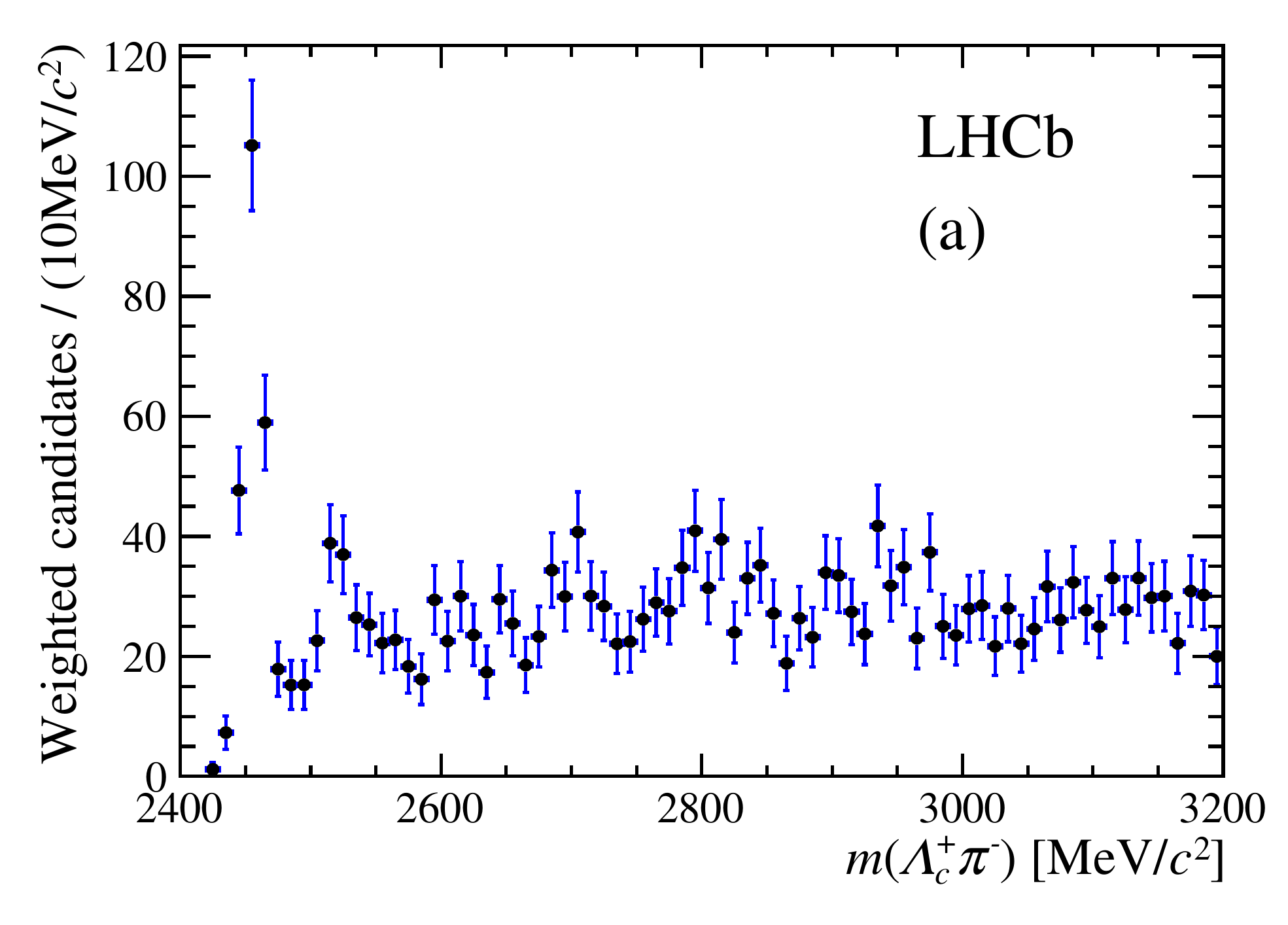}%
    \includegraphics[width=0.49\textwidth]{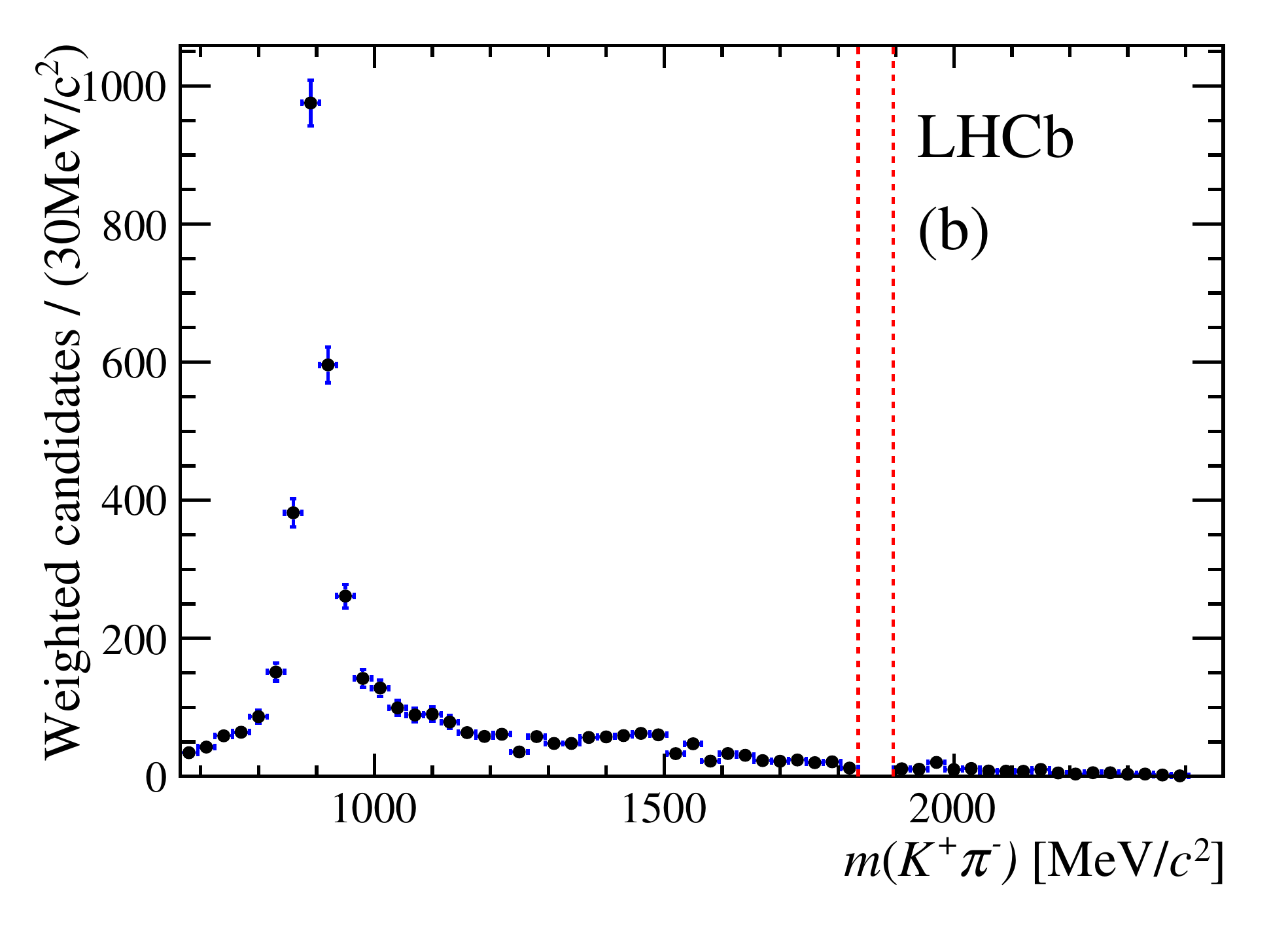}\\
    \includegraphics[width=0.49\textwidth]{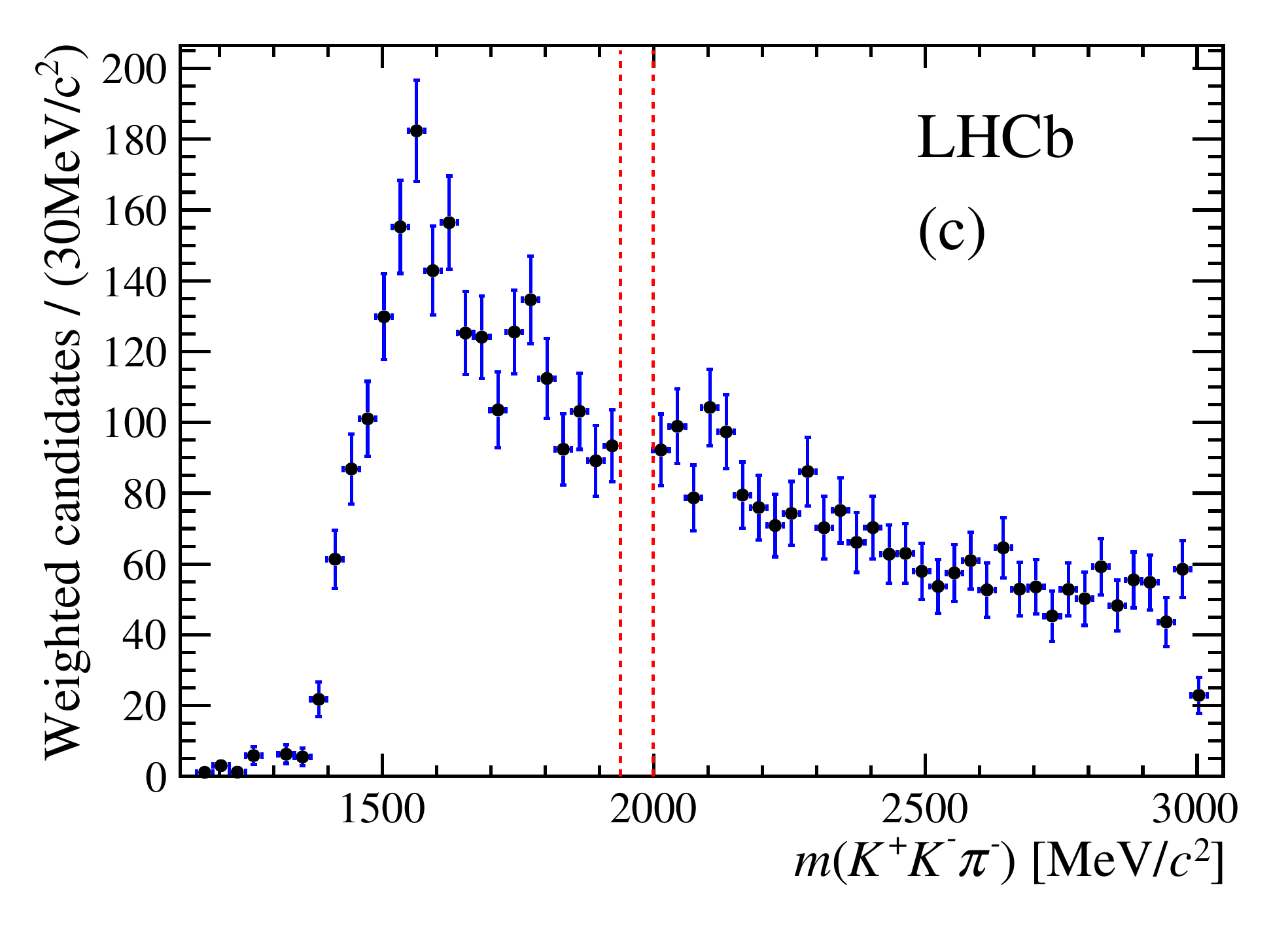}
    \vspace*{-0.5cm}
  \end{center}
  \caption{
    \small 
    Invariant mass distributions of (a) the \Lc\pim, (b) \Kp\pim, and (c) \Kp\Km\pim systems in the \LbLckkpi signal channel, 
    for the background-subtracted data. The red dashed vertical lines indicate the veto mass intervals for $\Dzb$ mesons in the $\Kp\pim$ distribution, and $\Dsm$ in the $\Kp\Km\pim$ distribution.
    }   
   \label{fig:res}
\end{figure}

Instead, a rich structure of known hadron contributions is visible in the background-subtracted invariant mass distributions of the $\Lc\pim$, $\Kp\pim$ and $\Kp\Km\pim$ systems, shown in Fig.~\ref{fig:res}. The $\PSigma_c(2455)^0$ and $\PSigma_c(2520)^0$ resonances are visible in the $\Lc\pim$ distribution. A large $K^*(892)^0$ resonance is observed in the $\Kp\pim$ projection. In the $\Kp\Km\pim$ system, a broad peaking structure at about $1.5$\gevcc is also observed. A similar structure is also seen in $\Bb\to D^{(*)}K^{*0}K^-$ decays by the Belle experiment~\cite{Drutskoy:2002ib}, and is explained as the tail contribution of the $a_1(1260)^-$ resonance.  

\section{Branching fraction ratio and efficiencies}
\label{sec:efficiency}

The ratio of the branching fractions of the \LbLckkpi decay including resonance contributions with respect to the normalisation channel is  determined by
\begin{eqnarray}
    \frac{\BR( \LbLckkpi)}{\BR( \LbLcDs)} & = & 
    \frac{N(\LbLckkpi)}{N(\Lb\to\Lc\Dsm[\Kp\Km\pim])} \times
    \label{eq:master} \\ 
    & & \frac{\epsilon_{\rm tot}(\Lb\to\Lc\Dsm[\Kp\Km\pim])}{\epsilon_{\rm tot}(\LbLckkpi)}
    \times \BR(\decay{\Dsm}{\Kp\Km\pim}), \nonumber
\end{eqnarray}
where \BR stands for the branching fraction of the corresponding decay.
The signal and normalisation yields are reported in Sec.~\ref{sec:fit}. The total efficiencies $\epsilon_{\rm tot}$ of the signal and the normalisation channels are determined by the product
\begin{equation}
\label{equ_eff}
\epsilon_{\rm tot} =\epsilon_{\rm acc}
\times \epsilon_{\rm sel} 
\times \epsilon_{\rm L0} 
\times \epsilon_{\rm HLT}
\times \epsilon_{\rm PID}, 
\end{equation}
where $\epsilon_{\rm acc}$ accounts for the \lhcb geometrical acceptance,
$\epsilon_{\rm sel}$ is the efficiency of reconstructing and selecting a candidate within the acceptance, 
$\epsilon_{\rm L0}$ is the L0 trigger efficiency for the selected candidates,
$\epsilon_{\rm HLT}$ is the HLT efficiency for the selected candidates passing the L0 trigger requirement,  and $\epsilon_{\rm PID}$ is the particle-identification (PID) efficiency for the selected candidates that survive all trigger requirements.
All efficiencies except for $\epsilon_{\rm\lone}$ and $\epsilon_{\rm PID}$ 
are determined from simulation, and the $(\pt,~y)$ distributions of the simulated \Lb baryons are weighted to match that of data, where $y$ is the rapidity of the candidate. 
The weights are obtained using the normalisation channel and applied to the signal decay. 

To take into account the resonance contributions to the signal decay channel, the simulation uses a mixture of three decay modes:  $\Lb\to\Lc a_1(1260)^-(\to K^{*0} K^-)$, $\Lc K^{*0} K^-$ and non-resonant four-body phase space. The fractions are determined by fitting the two-dimensional data distribution of $\Kp\pim$ and $\Kp\Km\pim$ invariant masses. 

The L0 efficiency of each hadron is computed using samples of well identified pions and kaons from $\Dz \to \Km \pip$ decays and protons from $\Lz \to p \pim$ decays~\cite{MartinSanchez:1407893}.
The efficiency is calculated in bins of transverse energy
for the particles incident on the \hcal surface, separately for its inner and outer regions. 
The PID efficiency is determined by the calibration samples of  $\Dstarp\to\Dz(\to\Km\pip)\pip$ and $\Lz\to\proton\pim$ decays and is evaluated as a function of track momentum, track pseudorapidity and event multiplicity, 
where the latter is represented by the number of the reconstructed tracks in the event.

The ratio between the total efficiencies for the signal and normalisation channels in Eq.~(\ref{eq:master}), is determined to be $0.78\pm0.02$, 
where the uncertainty accounts only for the size of the simulation sample. The value differs from unity primarily due to different selection efficiencies on the MVA responses for the signal and normalisation channels. 

External inputs are used for the branching fractions $\BR(\decay{\Dsm}{\Kp\Km\pim})=(5.39\pm0.15)\times10^{-2}$~\cite{PDG2020} and ${\BR( \LbLcDs)=(1.10\pm0.10)\times10^{-2}}$~\cite{LHCb-PAPER-2014-002}. In the latter case, while the value is measured by the LHCb collaboration~\cite{LHCb-PAPER-2014-002}, its uncertainty is dominated by the branching fraction of $\Bzb\to \Dp\Dsm$ decays, and is essentially uncorrelated with the present measurement.


\section{Systematic uncertainties}
\label{sec:systematic}

\begin{table}[b]
\centering
\caption{Summary of systematic uncertainties on the ratio of branching fractions.}
\label{tab:systemerr}
\begin{tabular}{l c }\hline\hline
    Source                                            &  Uncertainty (\%)  \\\hline
    Combinatorial background                                 &  0.9      \\
    Shape of $\Lb\to\Sigmacp\Kp\Km\pim$ contribution      &  0.3   \\
    \LbLckkpi background in normalisation channel        &  0.8   \\
    Signal fit model                                     &  0.5  \\
    Simulation sample size                                        &  2.5  \\
    PID efficiency                                       & 0.4   \\
    Trigger efficiency                                   & 0.1        \\
    $(\pt, ~y)$ weight                               & 0.8   \\
    Track multiplicity weight                          & 0.8   \\
    \Lc Dalitz structure                            & 1.4   \\
    Mixture fraction in simulation                        & 0.2   \\
    Resonance structure           & 3.6\\
    Multiple candidates                                  & 0.3\\
    MVA selection                                        & 0.5 \\ \hline

    Total& 4.9   \\ \hline \hline
\end{tabular}
\end{table}

All systematic uncertainties on the measurement of the ratio of branching fractions are listed in Table~\ref{tab:systemerr}.
The total uncertainty is determined from the sum  of all contributions in quadrature.
The dominant uncertainty is related to the resonance structure that is not perfectly modeled by the simulation.

Uncertainties due to the fit model are considered. For the background due to random combinations of final-state particles in both the signal and normalisation channels, the exponential function is replaced by a second-order polynomial function. From the comparison to the default result, the relative uncertainty on the ratio of branching fractions is 0.9\%.
In the signal channel, the uncertainty due to the $\Lb\to\Sigmacp\Kp\Km\pim$ background contribution is assessed
by performing the fit with a widened mass region, resulting in a relative uncertainty of 0.3\%.
For the normalisation channel, changing the yield of the \LbLckkpi contribution within its uncertainty results in a relative $0.8\%$ variation.

The systematic uncertainty due to the model for both signal and normalisation channels,
is studied by changing to a single Hypatia function~\cite{Santos:2013gra},
where the mean and width parameters are left free while all other parameters are taken from simulation.
This results in a relative uncertainty of $0.5\%$.

The uncertainties on the ratio of efficiencies are evaluated.
The uncertainty due to the finite simulation sample size is evaluated from the expected efficiency variation in bins of \pt and $y$ of the \Lb candidate as
\begin{equation}
    \sigma_{\epsilon} = \left.\sqrt{\sum_i {\epsilon_i(1-\epsilon_i) N_i w_i} } \middle/ 
    \sum_i N_i w_i\right.,
\end{equation}
for each bin $i$, where $N_i$ is the number of generated events, $w_i$ is a correction weight, and $\epsilon_i$ is the candidate efficiency. The normalisation of the weights is chosen such that the denominator is equal to total number of generated events without the weighting. The relative uncertainty is found to be $2.5\%$.

Pseudoexperiments are used to evaluate the systematic effects due to uncertainties on the weights or efficiencies in different bins. For a given source, many pseudoexperiments are generated, in which each produces a new set of weights or efficiencies according to the central values and uncertainties following Gaussian distributions. The efficiency ratio between the signal and normalisation channels is recomputed. The resulting efficiency ratios from many pseudoexperiments of this source produce a Gaussian distribution centering at the baseline value. The standard deviation of the Gaussian distribution is taken as absolute uncertainty on the efficiency ratio for the given source.
The procedure is applied to obtain the systematic uncertainty related to the PID and trigger efficiencies and to $(\pt,~y)$ and track multiplicity weighting.

The tracking efficiency returned by the simulation is calibrated using a data-driven method~\cite{LHCb-DP-2013-002}. The uncertainty on the calibration sample size is propagated to the efficiency ratio using pseudoexperiments, resulting in a systematic uncertainty of 0.8\%.   Because the final states for signal and normalisation modes are identical, possible data-simulation differences in hadron interactions with the detector material are assumed to be negligible.

The agreement between data and simulation for the $\Lc\to\proton\Km\pip$ channel is tested by comparing the Dalitz structure.
The signal simulation sample is weighted in the $m(\proton\Km)$
versus $m(\Km\pip)$ plane to match the distribution of the background-subtracted data.
The uncertainty related to the limited sample size used for obtaining these weights is 1.1\%, obtained from pseudoexperiments.
The uncertainty related to the choice of binning is 0.8\%, determined by using an alternative binning.
A total of 1.4\% is assigned as systematic uncertainty.

The contributions of the \Lb decays through the mixture of the three decay modes are considered when generating the simulated events of the signal channel,
and their fractions are obtained by fitting the two-dimensional
distribution of the \Kp\Km\pim and \Kp\pim systems in the background-subtracted signal data.
The fractions are changed according to the statistical uncertainty of the fit result,
yielding 0.2\% of relative uncertainty.

The simulation does not fully model the resonance structure, \eg the contribution of
 $\PSigma_c^{0}$ resonances, which  is clearly seen in the \Lc\pim invariant mass distribution, as illustrated in Fig.~\ref{fig:res}.
By weighting the simulation to match the $m(\Lc \pim)$ distribution in the data, a 1.3\% variation of the ratio of branching fractions is found and assigned as systematic uncertainty. Besides, differences between background-subtracted data and simulated signal events are also observed in the invariant mass distributions of the \Lc\Kp\Km and \Kp\Km systems.
To account for this discrepancy,
the simulated sample is weighted according to the \Lc\Kp\Km
or \Kp\Km mass distribution of background-subtracted data,
and the ratio of branching fractions is reevaluated. The two procedures return changes of 2.6\% and 2.0\%, respectively. The three values are added in quadrature to account for the uncertainty due to resonance structure.

Simulation does not account well for multiple candidates, which is found to be about 0.6\% of the data sample in the signal channel. Half of this fraction is assigned as systematic uncertainty due to the random choice to retain only one candidate.

The MVA selection criteria are optimized separately for the signal and normalisation channels.
As an alternative choice, the MVA selection of the normalisation channel is fixed to be the same as that of the signal channel to test the robustness of the MVA selection. The relative variation of the branching fraction ratio is 0.5\%, which is assigned as systematic uncertainty.


\section{Results and summary}
\label{sec:summary}

The first observation of the \LbLckkpi decay is presented, 
and the branching fraction is determined using the \LbLcDs decay as a normalisation channel.
The relative branching fraction is measured to be
\begin{equation*}
        \frac{\BR( \LbLckkpi)}{\BR( \LbLcDs)}=  (9.26 \pm 0.29 \pm 0.46 \pm 0.26)\times10^{-2},
\end{equation*}
where the first uncertainty is statistical, the second systematic, and the third is due to the knowledge of the \Dskkpi branching fraction~\cite{PDG2020}. Using this ratio, the \LbLckkpi branching fraction is determined to be  
\begin{equation*}
        \BR( \LbLckkpi)=  (1.02 \pm 0.03 \pm 0.05 \pm 0.10)\times10^{-3},
\end{equation*}
where the third term includes the uncertainty on the 
branching fraction of the $\LbLcDs$ decay~\cite{PDG2020}. The invariant mass distribution of the $\Lc\Kp$ system is inspected for possible structure due to  open-charm pentaquarks, and no contribution is observed.


\section*{Acknowledgements}
%
%
\noindent We express our gratitude to our colleagues in the CERN
accelerator departments for the excellent performance of the LHC. We
thank the technical and administrative staff at the LHCb
institutes.
We acknowledge support from CERN and from the national agencies:
CAPES, CNPq, FAPERJ and FINEP (Brazil); 
MOST and NSFC (China); 
CNRS/IN2P3 (France); 
BMBF, DFG and MPG (Germany); 
INFN (Italy); 
NWO (Netherlands); 
MNiSW and NCN (Poland); 
MEN/IFA (Romania); 
MSHE (Russia); 
MICINN (Spain); 
SNSF and SER (Switzerland); 
NASU (Ukraine); 
STFC (United Kingdom); 
DOE NP and NSF (USA).
We acknowledge the computing resources that are provided by CERN, IN2P3
(France), KIT and DESY (Germany), INFN (Italy), SURF (Netherlands),
PIC (Spain), GridPP (United Kingdom), RRCKI and Yandex
LLC (Russia), CSCS (Switzerland), IFIN-HH (Romania), CBPF (Brazil),
PL-GRID (Poland) and OSC (USA).
We are indebted to the communities behind the multiple open-source
software packages on which we depend.
Individual groups or members have received support from
AvH Foundation (Germany);
EPLANET, Marie Sk\l{}odowska-Curie Actions and ERC (European Union);
A*MIDEX, ANR, Labex P2IO and OCEVU, and R\'{e}gion Auvergne-Rh\^{o}ne-Alpes (France);
Key Research Program of Frontier Sciences of CAS, CAS PIFI,
Thousand Talents Program, and Sci. \& Tech. Program of Guangzhou (China);
RFBR, RSF and Yandex LLC (Russia);
GVA, XuntaGal and GENCAT (Spain);
the Royal Society
and the Leverhulme Trust (United Kingdom).


\addcontentsline{toc}{section}{References}

\bibliographystyle{LHCb}
\bibliography{main,standard,LHCb-PAPER,LHCb-CONF,LHCb-DP,LHCb-TDR, myBib}
 
\newpage
\centerline
{\large\bf LHCb collaboration}
\begin
{flushleft}
\small
R.~Aaij$^{31}$,
C.~Abell{\'a}n~Beteta$^{49}$,
T.~Ackernley$^{59}$,
B.~Adeva$^{45}$,
M.~Adinolfi$^{53}$,
H.~Afsharnia$^{9}$,
C.A.~Aidala$^{84}$,
S.~Aiola$^{25}$,
Z.~Ajaltouni$^{9}$,
S.~Akar$^{64}$
% ...

{\footnotesize \it

% ...
$ ^{9}$Universit{\'e} Clermont Auvergne, CNRS/IN2P3, LPC, Clermont-Ferrand, France\\
% ...
$ ^{25}$INFN Sezione di Milano, Milano, Italy\\
% ...
$ ^{31}$Nikhef National Institute for Subatomic Physics, Amsterdam, Netherlands\\
% ...
$ ^{45}$Instituto Galego de F{\'\i}sica de Altas Enerx{\'\i}as (IGFAE), Universidade de Santiago de Compostela, Santiago de Compostela, Spain\\
% ...
$ ^{49}$Physik-Institut, Universit{\"a}t Z{\"u}rich, Z{\"u}rich, Switzerland\\
% ...
$ ^{53}$H.H. Wills Physics Laboratory, University of Bristol, Bristol, United Kingdom\\
% ...
$ ^{59}$Oliver Lodge Laboratory, University of Liverpool, Liverpool, United Kingdom\\
% ...
$ ^{64}$University of Cincinnati, Cincinnati, OH, United States\\
% ...
$ ^{67}$Syracuse University, Syracuse, NY, United States\\
% ...
$ ^{84}$University of Michigan, Ann Arbor, United States, associated to $^{67}$\\
\bigskip
% ...
\medskip
}
\end{flushleft}


\end{document}